\documentclass[reprint,preprintnumbers,twocolumn,
superscriptaddress,
amsmath,amssymb,aps]{revtex4-2}
\usepackage[utf8]{inputenc}

\usepackage{graphicx}
\graphicspath{{./figs/}}
\usepackage{dcolumn}
\usepackage{bm}

\usepackage{mathrsfs}

\usepackage[normalem]{ulem}

\usepackage{color}
\usepackage[dvipsnames, table]{xcolor}
\definecolor{blue}{rgb}{0,0,0.5}
\definecolor{lightgray}{gray}{0.95} 

\usepackage[colorlinks,linkcolor=blue,urlcolor=blue,citecolor=blue]{hyperref}
\usepackage{graphicx}
\usepackage{slashed}
\usepackage{enumitem}
\usepackage{bm}
\usepackage{physics}
\usepackage{xspace}
\usepackage[T1]{fontenc}
\usepackage{lmodern}
   
\usepackage{tabularx} 
\usepackage{caption}  
\usepackage{booktabs}  
\usepackage{diagbox}
\usepackage{multirow}

\newcommand{\be}{\begin{equation}}
\newcommand{\ee}{\end{equation}}
\newcommand{\bea}{\begin{eqnarray}}
\newcommand{\eea}{\end{eqnarray}}

\newcommand{\mc}{\mathcal}

\newcommand{\nn}{\nonumber}
\newcommand{\noi}{\noindent}
\def\bk{\vb*{k}}

\def\bkV{\vb*{k}_{V}}
\def\bkA{\vb*{k}_{A}}
\def\hkL#1{\hat{\bk}_{L}{#1}}
\def\hkR#1{\hat{\bk}_{R}{#1}}

\newcommand{\MeV}{{\rm MeV}}
\newcommand{\GeV}{{\rm GeV}}
\def\refeq#1{Eq.~(\ref{#1})}
\def\reftab#1{Table~\ref{#1}}
\def\reffig#1{Fig.~\ref{#1}}

\def\la{\lambda}

\def\de{\partial}
\def\diag{{\rm diag}}
\def\DS#1{|\Delta S| = #1}

\newcommand{\df}{\mathscr{f}} 

\newcommand{\choicontact}{Choi:2021ign}
\newcommand{\choiRGE}{Choi:2021kuy}

\newcommand{\neubertPRL}{Bauer:2021wjo}
\newcommand{\neubertALPslong}{Bauer:2021mvw}
\newcommand{\scherer}{Scherer:2002tk}
\newcommand{\georgi}{Georgi:1986df}

\newcommand{\carenza}{Carenza:2020cis}

\newcommand{\mcJ}{\mc J}
\renewcommand{\df}{\mathcal{F}}
\newcommand{\beal}{\begin{aligned}}
\newcommand{\eeal}{\end{aligned}}
\newcommand{\sat}{{\rm sat}}
\newcommand{\cm}{{\rm cm}}

\newcommand{\Bimba}{B_i\,M\to B_f\,a}
\newcommand{\Biba}{B_i\,\to B_f\,a}
\newcommand{\DDmod}{{\tt DD2Y}\xspace}
\newcommand{\SFmod}{{\tt SFHoY}\xspace}

\makeatletter
\g@addto@macro\bfseries{\boldmath}
\makeatother

\makeatletter
\def\p@subsection{}
\makeatother

\newlist{todolist}{itemize}{2}
\setlist[todolist]{label=$\square$}
\usepackage{pifont}
\DeclareOldFontCommand{\rm}{\normalfont\rmfamily}{\mathrm}
\DeclareOldFontCommand{\sf}{\normalfont\sffamily}{\mathsf}
\DeclareOldFontCommand{\tt}{\normalfont\ttfamily}{\mathtt}
\DeclareOldFontCommand{\bf}{\normalfont\bfseries}{\mathbf}
\DeclareOldFontCommand{\it}{\normalfont\itshape}{\mathit}
\DeclareOldFontCommand{\sl}{\normalfont\slshape}{\@nomath\sl}
\DeclareOldFontCommand{\sc}{\normalfont\scshape}{\@nomath\sc}

\begin{document}

\preprint{LAPTH-006/24}
\preprint{DESY-24-011}

\title{Axion emission from strange matter in core-collapse SNe}

\author{Ma\"el Cavan-Piton}

\email{cavanpiton@lapth.cnrs.fr}

\affiliation{%
{\itshape LAPTh, Universit\'{e} Savoie Mont-Blanc et CNRS, 74941 Annecy, France}
}%

\author{Diego Guadagnoli}

\email{diego.guadagnoli@lapth.cnrs.fr}

\affiliation{%
{\itshape LAPTh, Universit\'{e} Savoie Mont-Blanc et CNRS, 74941 Annecy, France}
}%

\author{Micaela Oertel}

\email{micaela.oertel@obspm.fr}

\affiliation{%
{\itshape Laboratoire Univers et Th\'eories, Observatoire de Paris, Universit\'e PSL, CNRS, Universit\'e de Paris-Cité, 92190 Meudon, France}
}%

\author{Hyeonseok Seong}

\email{hyeonseok.seong@desy.de}

\affiliation{%
{\itshape Deutsches Elektronen-Synchrotron DESY, Notkestr. 85, 22607 Hamburg, Germany}
}%

\author{Ludovico Vittorio}

\email{ludovico.vittorio@lapth.cnrs.fr}

\affiliation{%
{\itshape LAPTh, Universit\'{e} Savoie Mont-Blanc et CNRS, 74941 Annecy, France}
}%

\begin{abstract}
\noi The duration of the neutrino burst from the supernova event SN~1987A is known to be sensitive to exotic sources of cooling, such as axions radiated from the dense and hot hadronic matter thought to constitute the inner core of the supernova.
We perform the first quantitative study of the role of hadronic matter beyond the first generation---in particular strange matter. We do so by consistently including the full baryon and meson octets, and computing axion emissivity induced from baryon-meson to baryon-axion scatterings as well as from baryon decays. We consider a range of supernova thermodynamic conditions, as well as equation-of-state models with different strangeness content. We obtain the first bound on the axial axion-strange-strange coupling, as well as the strongest existing bound on the axion-down-strange counterpart. Our bound on the latter coupling can be as small as $O(10^{-2})$ for $f_a = 10^9~\GeV$.
\end{abstract}

\maketitle

\noi {\bf Introduction---} Axions are among the best motivated and at present most sought-after particles beyond the Standard Model (SM). Searches and predictions follow two streams of activity. The first concerns the ``invisible'' \cite{Kim:1979if,Shifman:1979if,Dine:1981rt,Zhitnitsky:1980tq} QCD axion \cite{Peccei:1977hh,Peccei:1977ur,Weinberg:1977ma,Wilczek:1977pj}, whose mass $m_a$ and decay constant $f_a$ are related---the smaller $m_a$ the feebler the axion's interaction strength with known matter. The second stream of activity is the more general framework of axion-like particles (ALPs), where $m_a$ and $f_a$ are considered as independent parameters, constrained mostly by data (for UV aspects, see \cite{Svrcek:2006yi,Arvanitaki:2009fg}).

In the QCD-axion case already, couplings to matter are very model-dependent and are constrained through a wide and growing  array of laboratory \cite{Irastorza:2018dyq,Sikivie:2020zpn}, astrophysical \cite{Raffelt:1990yz,Raffelt:2006cw,Caputo:2024oqc} or cosmological \cite{Marsh:2015xka} observables, as well as by theoretical considerations \cite{Kim:2008hd,GrillidiCortona:2015jxo,DiLuzio:2020wdo,Choi:2020rgn}. Among astrophysical observables, core-collapse supernov{\ae} (SNe) such as SN~1987A stand out. Measurements of the neutrino burst associated to SN~1987A 
\cite{Bionta:1987qt,IMB:1988suc,Kamiokande-II:1987idp,Hirata:1988ad,Alekseev:1987ej,Alekseev:1988gp}
 match expectations~\cite{Burrows:2000mk,Woosley:2005cha}, and this provides a strong constraint on non-standard sources of cooling such as free-streaming axions $a$ coupled to nucleons~\cite{Turner:1987by,Raffelt:1987yt,Burrows:1988ah}. This is usually quantified through $Q_a < Q_\nu$~\cite{Raffelt:1990yz,Burrows:1988ah}, with $Q$ denoting the power radiated in the given particle per unit volume, or emissivity. The understanding of the different processes contributing to $Q_a$ has seen rapid progress over recent years. The most established bound comes from axion-strahlung in nucleon-nucleon scattering \cite{Carenza:2019pxu}, although a reliable calculation is challenging (see discussion in Ref.~\cite{Caputo:2024oqc}). One class of processes that has received particular attention are pion-nucleon reactions $\pi N \to N^\prime a$, which may even be dominant \cite{Carenza:2020cis,Fischer:2021jfm,Choi:2021ign,Fore:2019wib,Lella:2022uwi,Ho:2022oaw,Vonk:2022tho,Lella:2023bfb}. A further process considered recently is $\Lambda \to n \, a$, see~\cite{MartinCamalich:2020dfe,Camalich:2020wac,DEramo:2021usm}.

The physics of axion emission from SNe is, however, still far from established in numerous respects, among the others: the hydrodynamical modelling of the SN, consistently including axion emission at the dynamical level as well as possible departures from spherical symmetry, see in particular \cite{Fischer:2021jfm,Mori:2021pcv,Betranhandy:2022bvr,Mori:2023mjw};
the question how in-medium effects impact the leading axion-emission processes; the possible role of axion interactions with matter components inside the SN volume {\em beyond} the first generation of matter. Our work aims to perform a benchmark calculation that quantitatively addresses this last aspect. To this end, we evaluate the contributions to axion emissivity from all processes of the kind $\Bimba$, where $B_{i,f}$ and $M$ denote octet baryons and mesons, respectively, as well as the decays $\Biba$. Their effect is non-trivial for the following reason. These contributions correspond to over one hundred different reactions, each  yielding a {\em positive}-definite contribution to the emissivity $Q_a$. (By this argument, any additional process not included in our calculation would only make our conclusions stronger.) This large number of processes compensates for the suppressed densities of strange baryons relative to those of ordinary nucleons~\cite{Oertel:2016xsn,daSilvaSchneider:2020ddu,Malik:2020jlb,Kochankovski:2022rid,Raduta:2022elz,Sedrakian:2022ata}. As a result, $Q_a$ serves as a non-trivial probe of axion interactions with strange matter. Specifically, the constraint from SN 1987A introduces new correlations between flavour-diagonal axial couplings involving an axion and any of the $u,d,s$ quarks. Besides, this constraint introduces a {\em new bound} on the magnitude of the flavor-violating counterpart. This bound spans $O(10^{-1}$-$10^{-2})$, the range being determined by a variation of the thermodynamic conditions inside the axion-emitting SN core within realistic domains. Our bound assumes $f_a = 10^9\, \GeV$, but can be scaled to a different $f_a$ choice through a well-defined procedure detailed later.
Our work is, to our knowledge, the first study of axion emission from beyond-first-generation matter in the consistent framework of an effective theory, in particular including all the allowed processes mentioned above.

\noi {\bf Setup---} We strived to test in detail the robustness of our conclusions with respect to the two main modelling aspects inherent in the system we consider: on the one hand, the modelling of the SN volume affected by axion emission, and on the other, of the axion-hadron interactions responsible for axion emission. We discuss these two key aspects in turn.

Our SN-core modelling is based on the following guidelines. Assuming weak equilibrium for the interactions that involve the strange quark, and neglecting the presence of muons, the state of matter is characterised by three thermodynamic parameters, chosen usually as the temperature $T$, baryon number density $n_B$ and electron fraction $Y_e \equiv (n_{e^-} - n_{e^+})/n_B$. These parameters allow to determine the abundances of all particles entering the equation of state~\cite{Janka:2006fh,Oertel:2016bki}.
These parameters are, in general, {\em local} within the SN volume, and their values are obtained by solving the hydrodynamic equations from energy-momentum, baryon and electronic lepton number conservation, coupled to neutrino transport, see e.g.~\cite{Bruenn:1985en}.
Inclusion of axion transport accordingly allows to account consistently for energy and momentum loss through axion emission, as described e.g. in \cite{Fischer:2021jfm,Mori:2021pcv,Betranhandy:2022bvr,Mori:2023mjw}.
We do not perform here such a full-fledged SN simulation.
Instead, we assume homogeneous thermodynamic parameters within the SN axion-sphere, in an approach similar to much of the literature on axion emission from SNe \footnote{This picture---which builds on the fact that the main observable is the duration of the $\nu$ signal, whereas the amount of energy released is an inferred quantity---lends itself to the use of emissivity \cite{Raffelt:1990yz} to place the constraint, as we will discuss around \refeq{eq:Qa}.}.
For the parameters we choose sets of values that should be representative of realistic conditions: $T = \{30, 40\}~\MeV$; $n_B = \{ 1, 1.5 \} \, n_{\sat}$, with $n_{\sat} = 1.6 \times 10^{38} \, \cm^{-3}$. As the third thermodynamic parameter, we take the total baryonic e.m. charge fraction $Y_{Q_B} \equiv \sum_{i \in B} Q_i n_i/n_B = 0.3$. Bearing in mind that hadron number fractions are dominated by nucleons (any other hadron contributing at the percent level or less) $Y_{Q_B} = 0.3$ ensures $Y_p \simeq 0.3$, consistent with the value adopted in e.g. \cite{\carenza}.
We note that the emissivity varies strongly with temperature---higher temperatures yielding larger $Q_a$ values.
We consider $T = 30~\MeV$ as it complies with a standard choice in the literature, and it also provides somewhat more conservative bounds from $Q_a$ with respect to a higher $T$ value. The case $T = 40~\MeV$ is, in turn, useful to quantify the effect of temperature variation in a reasonable, yet large enough range.

On top of the above thermodynamic-parameter choices, we also consider two equation-of-state (EoS) models consistently containing the full baryonic octet, \DDmod \cite{Marques:2017zju} and \SFmod \cite{Fortin:2017dsj}. This choice is guided by the following arguments among the others: these models specifically predict different amounts of strangeness inside hot and dense matter; they are publicly available; they are compatible with astro- and nuclear-physics constraints \cite{Marques:2017zju,Fortin:2017dsj}. Hence, calculating $Q_a$ within these two models provides a direct way to establish how sensitively our conclusions depend on the assumed amount of strange matter in the SN core.
The meson octet is treated as an ideal Bose gas with chemical potentials obtained from the EoS model.
Including an interaction between mesons and baryons would slightly modify the particle fractions, but we expect this to have a negligible impact on our conclusions~\footnote{We note that the effect appears to be small at the temperatures of interest in our case \cite{Fore:2019wib}.}.
The $e^-$ fraction is then obtained from the condition of charge neutrality $Y_{Q_B} + Y_{Q_M} + Y_e = 0$. 

\begin{figure}[t]
  \begin{center}
  \includegraphics[width=0.20\textwidth]{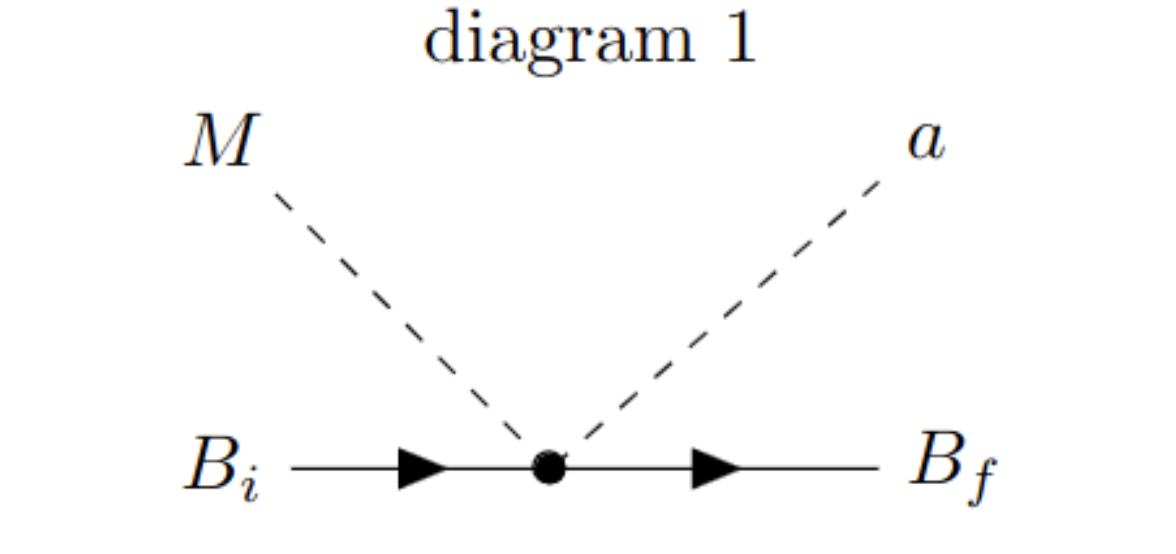} \hspace{0.2cm}
  \includegraphics[width=0.20\textwidth]{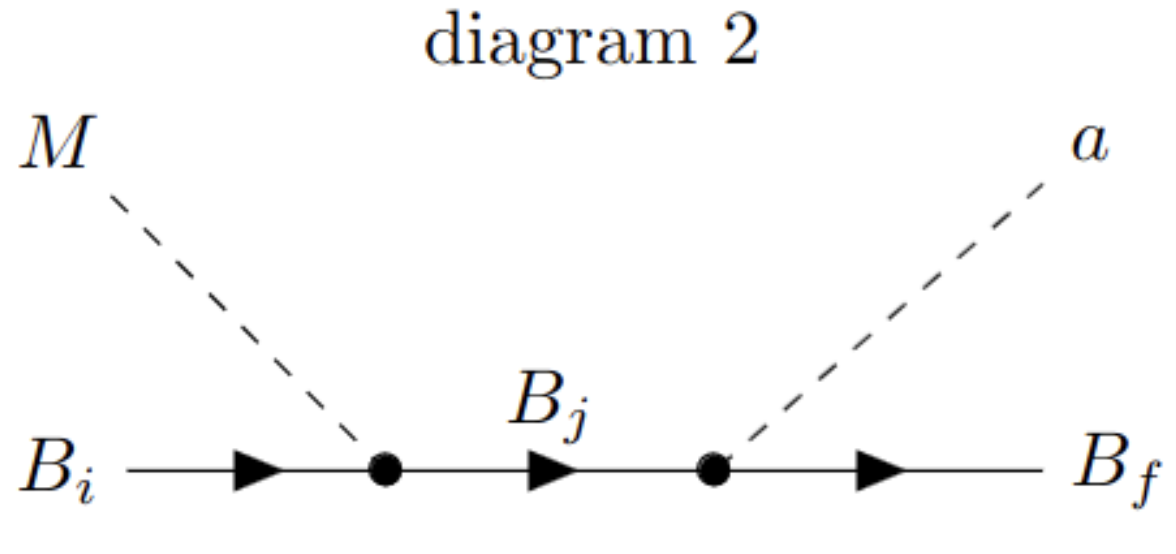} \vspace{0.2cm}

  \includegraphics[width=0.20\textwidth]{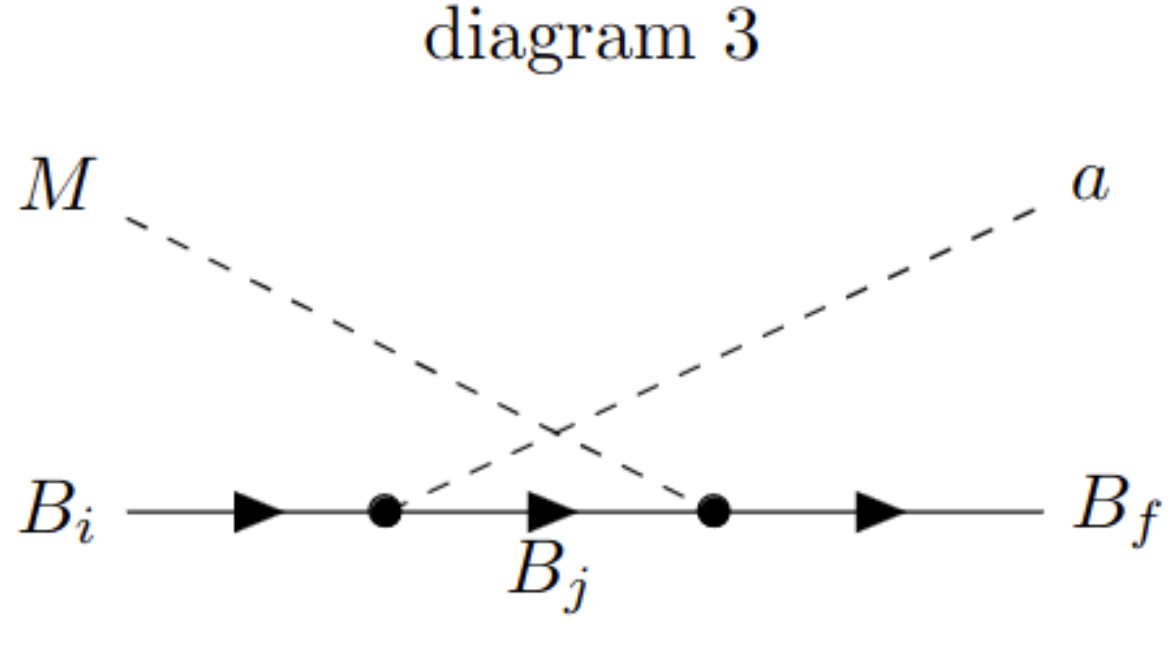} \hspace{0.2cm}
  \includegraphics[width=0.20\textwidth]{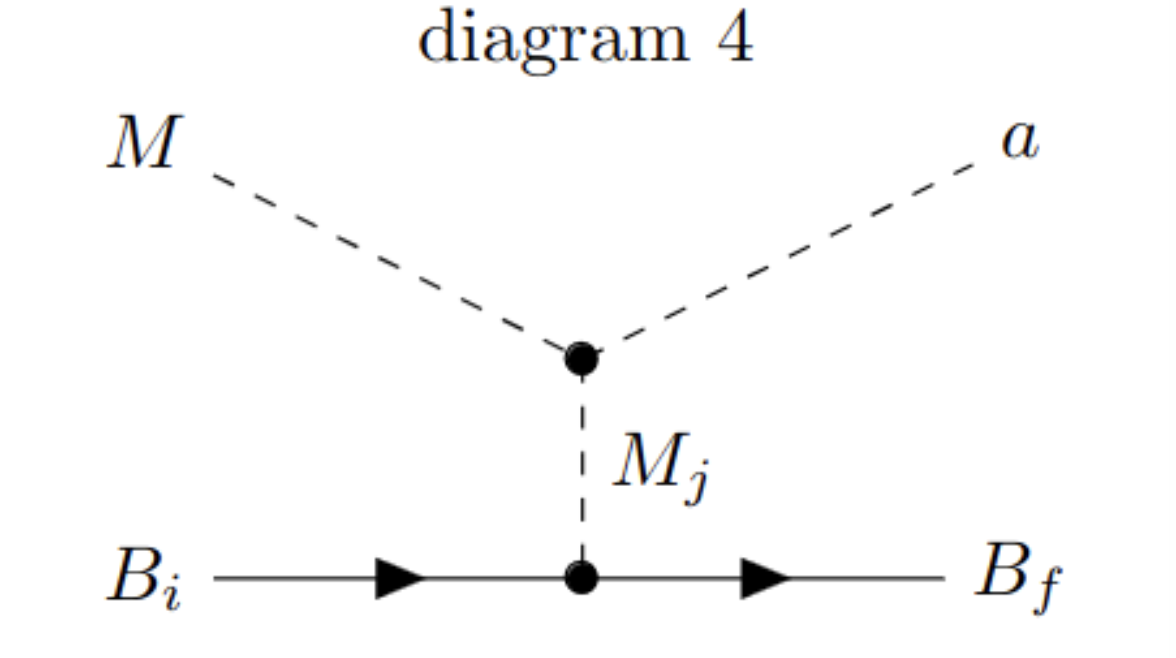}
  \end{center}
  \caption{The diagrams contributing to $\Bimba$, with $B_{i,f}$ initial- or final-state octet baryons, $M$ octet mesons, and $a$ the axion.
\label{fig:BiM}
}
\vspace{-0.5cm}
\end{figure}

The two EoS models, applied on the $T$ and the $n_B$ values discussed above, represent our ensemble of thermodynamic conditions. Specifically, each of them translates into a set of values for the chemical potentials of charged and neutral octet baryons and mesons. This allows to calculate the axion emissivity $Q_a$ for all the baryon-meson reactions of the type $\Bimba$, as well as the decays $\Biba$. The possible processes are listed in Sec.~\ref{app:processes} of the Supplemental Material~\footnote{See Supplemental Material for a detailed list of allowed processes; explicit formulae for axion-hadron interactions, also to fix notation; technical details on Fig. 2, and comparison plots within the SFHoY model; details on, and a listing of, SN particle fractions within our considered SN models. The Supplemental Material also makes reference to Refs.~\cite{\scherer,Vonk:2021sit,Cronin:1967jq,Kambor:1989tz,Notari:2022ffe,Typel:2013rza,CompOSECoreTeam:2022ddl}.}.
The diagrams relevant to $\Bimba$ are depicted in \reffig{fig:BiM}.

The second modelling aspect that we need to specify is that of the axion-hadron interactions. The latter may be obtained through the Noether procedure applied to the 3-flavour octet-mesonic as well as -baryonic chiral Lagrangians, augmented with an axion as elucidated in~\cite{Georgi:1986df}. Denoting the octet-meson and octet-baryon fields as $U$ and $B$ respectively, one gets
\be
\label{eq:LaUB}
\mc L_{a U B} ~=~ \frac{\de_\mu a}{f_a} \sum_{b =1}^8
\left( x^b_L~\mc J_L^{b \mu}(U,B) + x^b_R~\mc J_R^{b \mu}(U,B)
\right)~,
\ee
where $x^b_{L,R}(a) ~\equiv~ \Tr(\hat{\vb*{k}}_{L,R}(a) \la^b)$ are the axion-hadron coupling strengths of each of the interaction currents $\mc J_{L,R}^{b \mu}$ (collected for completeness in the supplemental Sec.~\ref{app:currents}). As is to be expected, these strengths are functions of $\hat{\vb*{k}}_{L,R}$, i.e. the fundamental axion-quark coupling matrices defined from the Lagrangian \cite{Bauer:2021wjo}
\be
\label{eq:Laqq}
\mc L_{aqq} ~\equiv~ \frac{\de_\mu a}{f_a} 
\left(
\bar q \, \gamma_L^\mu \hat{\vb*{k}}_{L}{(a)} \, q + 
\bar q \, \gamma_R^\mu \hat{\vb*{k}}_{R}{(a)} \, q
\right)~,
\ee
with $q = (u, d, s)^T$ and $\gamma_{L,R}^\mu = \gamma^\mu(1\mp\gamma_5)/2$.

As we consider processes involving a single axion, we can replace $\hat{\vb*{k}}_{L,R}(a) \to \vb*{k}_{L,R}$. Axion emissivity constrains the ratios $(\vb*{k}_{L,R})_{ij} / f_a$, or equivalently, the $\vb*{k}_{V,A}\equiv\vb*{k}_{R}\pm \vb*{k}_{L}$ counterparts. On the possible values for the indices, $ij = \{11, 22, 33, 23, 32\}$, we note the following.
Hermiticity requires the diagonal entries to be real and the off-diagonal ones to have opposite phases; e.m. charge conservation sets to zero off-diagonal entries other than 23, 32. We have therefore a total of 10 parameters from $\bk$ couplings, of which $(\vb*{k}_{V,A})_{23}$ are genuine beyond-SM sources of strangeness-changing currents. 

The $\bk$ couplings are free parameters \cite{\georgi}, fixed only by a theory of flavour, or by data.
We stick to an agnostic approach, making no theoretical assumptions about any of the couplings. We sample them using uniform distributions over ranges covering several orders of magnitude and constrain them based on measurements. This ensures that the bounds we obtain are as conservative as possible.
The entries $(\bk_{A})_{11}$ and $(\bk_{A})_{22}$ are bounded from data on isolated-Neutron-Star (NS) cooling, as recently discussed in Ref.~\cite{Buschmann:2021juv}. To maintain our conservative stance, we have considered the scenario where axion couplings to neutrons and protons are both non-zero, and $\lesssim 10^{-9}$, which saturates the NS cooling bound.
The couplings $(\vb*{k}_{V,A})_{23}$ are generally constrained from kaon processes (henceforth ``$K$ bounds''), in particular $K^0 - \bar K^0$ mixing and $\Gamma(K \to \pi a)$. The latter process is sensitive to $|(\bkV)_{23}|$ alone, while the $K^0 - \bar K^0$-mixing constraint from the $\epsilon_K$ observable \cite{ParticleDataGroup:2022pth} jointly sets bounds on $(\bkV)_{23}$ and $(\bkA)_{23}$, see Ref.~\cite{MartinCamalich:2020dfe}. We compare the theoretical predictions against the $\Gamma(K \to \pi a)$ bound~\cite{E949:2007xyy}, as well as the 95\% confidence-level range $0.88 < C_{\epsilon_K} < 1.36$ on $C_{\epsilon_K} \equiv |\epsilon_K^{{\rm SM}+a}| / |\epsilon_K^{\rm SM}|$, using the latest analysis by the UTfit collaboration \cite{UTfit_note}~\footnote{We note that this range is only available on~\cite{UTfit_note} as an online update, and is more recent that the latest published update in Ref. \cite{Bona:2022zhn}. Our used range is nearly identical to the one used in Ref.~\cite{MartinCamalich:2020dfe}}.

Some comments are in order on vectorial couplings. First, $Q_a$ is insensitive to diagonal $\bk_V$ couplings, that are solely observable via axion-hadron interactions whose underlying quark transition occurs via the weak force~\cite{\georgi,\neubertPRL,\neubertALPslong}. In addition, the $Q_a$ bound on $|(\bk_{V})_{23}|$ is less stringent than the one from the non-observation of $K \to \pi a$. Finally, the $Q_a$ bound is also insensitive to the (difference between the) $(\bk_{V})_{23}$ and $(\bk_{A})_{23}$ phases.

\noi {\bf Results---} We next discuss our numerical results. We calculate the axion emissivity $Q_a$ as~\cite{Tamborra:2017ubu}
\bea
\label{eq:Qa}
&&\hspace{-0.6cm}Q_a ~=~ \\
&&\hspace{-0.6cm}\int E_a ~ (2\pi)^4 \delta^4(p) \, \left\vert \mc M \right\vert^2 \df_i \, \df_M \, (1 - \df_f) \, \prod_{m}^{\{i,f,M,a\}} \frac{d^3 \vb*{p}_m}{(2\pi)^3 2 E_m}~,\nn
\eea
where $\left\vert \mc M \right\vert^2$ denotes the absolute value squared of the amplitude $\sum_n \mc A(\Bimba)$, $n = \{ i,f,M\}$,
where the indices $i, f, M$ label the different possible instances of initial-, final-state baryon, and meson involved in the reactions.
The distribution functions are defined as 
\be
\df_{j} ~=~ \frac{1}{\exp \left( \frac{E_{j} - \mu_{j}}{T} \right) + (-1)^{2 s_j + 1}}~,
\ee
where $s_j$ denotes the spin of the corresponding particle. \refeq{eq:Qa} generalizes to decays of the kind $\Biba$ in an obvious way.
As written, the emissivity treats all particles involved as ideal gases.
However, in-medium effects in hot and dense matter are expected to have a strong impact~\cite{Martinez-Pinedo:2012eaj,Roberts:2012um}. Such effects may be captured at the mean-field level by shifting effective potentials, masses and energies of all particles entering the EoS with their ``effective'' counterparts (e.g. Refs.~\cite{Martinez-Pinedo:2012eaj,Roberts:2012um,Reddy:1997yr,Oertel2020}). Our case of more dilute matter is expected to yield weaker mean-field effects, but we leave the investigation of this issue to a separate study. We thus perform the replacements
\be
m_j \to m_j^*~,~~E_j \to E_j^*~,~~\mu_j \to \mu^*_j = \mu_j - U_j~,
\ee
where the r.h.s.'s are calculated within each of our considered thermodynamic conditions.
In particular, the effective chemical potential contains the mean-field interaction potential $U_j$, which also enters the energy-conservation equation and is determined self-consistently.
This treatment of mean-field effects also allows to correctly recover the particles' number densities by integrating their distribution functions over phase space.

We made two further validations. A first one was to employ the setup in Ref.~\cite{Carenza:2020cis}, namely to restrict to the $\pi^- p \to n\,a$ process and further drop the first and fourth diagram in \reffig{fig:BiM} (omitted in Ref.~\cite{Carenza:2020cis}) \footnote{This comparison actually requires taking into account minor typos in Eqs. (5) and (11) of Ref.~\cite{Carenza:2020cis}. See also discussion in Ref.~\cite{Lella:2022uwi}.}. In such setup we find our results to be consistent with the predictions in Table~I of Ref.~\cite{Carenza:2020cis}.
A second check was to verify that the parameter bounds inferred from $Q_a$ yield an axion mean free path~\cite{Martinez-Pinedo:2012eaj} well above the SN radius, which validates the free-streaming assumption that underlies the $Q_a$ calculation~\cite{Raffelt:1990yz}.

We are now ready to quantitatively discuss the implications of the bound $Q_a < Q_{\nu}$~\cite{Raffelt:1990yz,Burrows:1988ah} within the different thermodynamic conditions and the different scenarios for the $\vb*{k}$ couplings considered. The coherence of the results we obtain suggests the following general picture.
\begin{figure*}[th!]
  \centering
  \includegraphics[width=\textwidth]{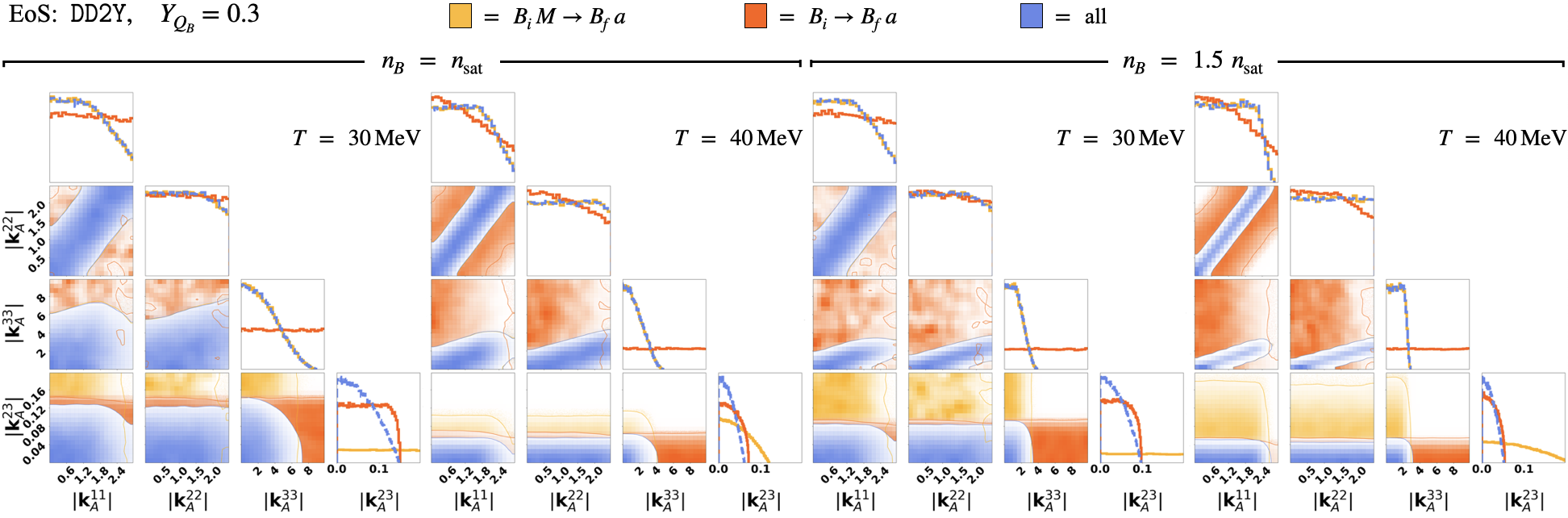}
  \caption{Regions allowed to $|(\bk_{A})_{ij}|$, $ij = \{11,22,33,23\}$ by $Q_a(\Bimba)$ (yellow), $Q_a(\Biba)$ (orange) and their sum (blue) $n_B = n_{\sat}$ (first two panels) or $n_B = 1.5 \, n_{\sat}$ (last two panels).
The plots on the diagonal of each panel display the posterior distribution for the respective $\bk$ coupling.
The contour lines in the plots off-the-diagonal display marginalized 2D 95\% posterior probabilities.
The figure uses the \DDmod EoS with $Y_{Q_B} = 0.3$.}
  \label{fig:matrix_plots_DD2Y_Sag}
\end{figure*}
Irrespective of the model, axion emissivity introduces non-trivial constraints on all of $|(\bk_{A})_{ij}|$, namely $ij = \{11,22,33,23\}$. In particular, {\em (a)}~$Q_a$ generates correlations between flavour-diagonal $\bk_A$ couplings, due to its proportionality to the square of well-defined linear combinations of $(\bk_{A})_{ii}$ entries. Also, $Q_a$ introduces a bound, albeit of $O(1)$ and thus mild, on all these entries; {\em (b)}~$Q_a$ introduces a novel, direct constraint on $|(\bk_A)_{23}|$. The bound can be as low as $10^{-2}$ (for our reference $f_a$ value), depending on the thermodynamic conditions chosen.

The above findings are illustrated in more detail in \reffig{fig:matrix_plots_DD2Y_Sag}, which captures all the main messages of this study, but requires a detailed look. Each panel is a ``matrix'', showing the correlation of every parameter ($|(\bk_{A})_{11,22,33,23}|$ in the $x$ axis) against every other one. The diagonal entries of the matrix are thus 1-parameter histograms normalized to unity, where the remaining parameters are marginalized over, i.e. assumed to take on every possible value.
The four instances of matrix plots in \reffig{fig:matrix_plots_DD2Y_Sag} refer to the two temperatures $T$ and the two baryonic densities $n_B$ considered, as specified in the legend.
The three colors show the $Q_a$ bound inferred from $\Biba$ processes only (red), from $\Bimba$ ones only (yellow), and from all (blue). The correlations (item {\em (a)}) are clearly visible as oblique blue bands; the bounds (item {\em (b)}) are the endpoints of the histograms along the diagonal of the matrix plots. As the lowermost histogram in each panel's diagonal shows, the $(\bk_A)_{23}$ bound we obtain is highly non-trivial, and gets as low as $5 \cdot 10^{-2}$ within the range of our considered thermodynamic conditions.
The figure also suggests that: {\em (i)}~correlations become sharper and bounds tighter if $T$ increases (first vs. second, or third vs. fourth panel); {\em (ii)}~the bound on $(\bk_A)_{33}$ is always dominated by $\Bimba$ processes (yellow constraints); {\em (iii)}~the bound on $(\bk_A)_{23}$ is dominated by $\Biba$ (red), although $\Bimba$ processes (yellow) increase in importance for larger $T$.
We also find that an increase in $n_B \in \{ 1, 1.5\} \, n_{\sat}$ generally sharpens both findings {\em (a)} and {\em (b)}. This is to be expected, as increasing $n_B$ generally increases the hyperon phase space.

\newcommand{\mystackrel}[2]{\stackrel{#1}{{\bf #2}}}

\begin{table}[t]
  \caption{$Q_a$ bounds on $|(\bk_A)_{23,33}|$, assuming $f_a = 10^9~\GeV$. 
  The larger boldfaced vs. smaller value quoted in each table entry refers to the EoS model being considered, \DDmod \cite{Marques:2017zju} vs. \SFmod \cite{Fortin:2017dsj} (see text for details on these models).
  }
  \centering
  \begin{tabularx}{\columnwidth}{>{\centering\arraybackslash}p{0.27\columnwidth} | >{\centering\arraybackslash}p{0.16\columnwidth} > {\centering\arraybackslash}p{0.16\columnwidth} | >{\centering\arraybackslash}p{0.16\columnwidth} >{\centering\arraybackslash}p{0.16\columnwidth}}
  \toprule 
{\vspace{-0.03cm}$\bk$ coupling}  
& \multicolumn{2}{c|}{\begin{tabular}[t]{@{}c@{}}{$n_B = n_{\sat}$}\vspace{0.1cm} \\ {$30$ MeV~~~~$40$ MeV}\end{tabular}} & \multicolumn{2}{c}{\begin{tabular}[t]{@{}c@{}}{$n_B = 1.5 \, n_{\sat}$}\vspace{0.1cm} \\ {$30$ MeV~~~~$40$ MeV}\end{tabular}} \\
  \midrule
  $|(\bk_A)_{23}|$ & $\mystackrel{0.35}{0.15}$ & $\mystackrel{0.12}{0.061}$ & $\mystackrel{0.38}{0.097}$ & $\mystackrel{0.14}{0.052}$ \\
  \addlinespace
  $|(\bk_A)_{33}|$ & $\mystackrel{8.8}{8.9}$ & $\mystackrel{4.4}{4.8}$ & $\mystackrel{5.9}{3.9}$ & $\mystackrel{3.1}{2.9}$ \\
  \bottomrule
  \end{tabularx}
  \label{tab:bounds}
\end{table}
The largest possible values shown in the blue histograms (i.e. the histograms corresponding to $Q_a$ calculated using all processes, $\Bimba$ as well as $\Biba$) on the diagonal of each panel of \reffig{fig:matrix_plots_DD2Y_Sag} are also tabulated in \reftab{tab:bounds}. More precisely, they correspond to the large, boldfaced entries in this table, which refer to the \DDmod EoS model. The smaller entries on top of the large ones refer to the \SFmod EoS model instead, and are reported for direct comparison. This comparison shows at a glance that the \DDmod model leads to stronger bounds than \SFmod.
Bounds are typically the result of several competing effects, and a numerical evaluation is necessary.
For example, hyperon densities tend to be higher within \DDmod than they are in \SFmod (see also supplemental Sec.~\ref{app:snmodel}). On the other hand, for reactions involving hyperons as final state, smaller fractions are even advantageous.

We emphasize that, in both \reffig{fig:matrix_plots_DD2Y_Sag} and \reftab{tab:bounds}, the $\vb*{k}$ couplings are constrained solely by data, as described in the text between Eqs.~(\ref{eq:Laqq}) and (\ref{eq:Qa}). In other words, we make no model assumptions regarding the $\vb*{k}$ couplings. It is useful to discuss how our obtained bounds would be modified if one imposes relations on the $\vb*{k}$ couplings inferred from a theory of flavour. We mention two representative examples. A first one is the minimal QCD-axion solution to the flavour problem~\cite{Ema:2016ops,Calibbi:2016hwq,Reiss:1982sq,Wilczek:1982rv,Ahn:2014gva,Bjorkeroth:2017tsz}. We find that $Q_a$ is more constraining than the theory assumptions within this model, and our bounds in \reftab{tab:bounds} remain valid.

A second example is Minimal Flavour Violation (MFV) \cite{Chivukula:1987py,Hall:1990ac,DAmbrosio:2002vsn,Buras:2000dm} within axion models \cite{Choi:2017gpf, Arias-Aragon:2017eww}. 
The minimal radiative flavour violation necessarily induces a very small left-handed off-diagonal component, scaling as $(\vb*{k}_{L})_{23} \sim \frac{1}{16\pi^2} (V_{\rm CKM})^*_{31} (V_{\rm CKM})_{32}$ multiplied by flavour-{\em universal} logs of ratios of ultraviolet scales. This functional dependence implies very suppressed values of  $O(10^{-6})$. Besides, the right-handed counterpart is absent by definition.
As a consequence $(\vb*{k}_{V,A})_{23} \approx  (\vb*{k}_{L})_{23} = O(10^{-6}) $ or smaller within this scenario.
In turn, the couplings $(\vb*{k}_{A})_{11,22,33}$ can be matched onto the axion-quark couplings $C_{u,d,s}$ of Refs.~\cite{\choicontact,\choiRGE}.
At a renormalization scale $\mu \approx 1\,\GeV$ one has $C_{u,d,s} = O(10^{-2})$ within KSVZ models \cite{Kim:1979if,Shifman:1979if}, and generically of $O(10^{-1})$ within DFSZ \cite{Dine:1981rt,Zhitnitsky:1980tq}. We thus see that, for any of the $\vb*{k}$ couplings, $Q_a$ is not constraining within this scenario, i.e. theory assumptions play a dominant role for all constrainable couplings.

We finally note that our results assume $f_a = 10^9~\GeV$, which is the reference value throughout our numerical study. We decided to keep $f_a$ fixed for two reasons. First, $f_a$ can vary by orders of magnitude, and such variation would completely obliterate the $Q_a$ dependence on $\vb*{k}$ couplings, which are our focus. The question of what is the limit on $f_a$ implied by our study necessarily requires to fix $\vb*{k}_{11,22,33}$ and $|\vb*{k}_{23}|$ first.
Second, the $Q_a$ dependence on $f_a$ is ruled by a simple scaling: if $f_a \to f_a / x_f$, $Q_a \to Q_a x_f^2$.

In conclusion, we have presented the first study of SN axion emissivity $Q_a$ that consistently quantifies the possible role of matter beyond the first generation, by calculating all baryon-meson to baryon-axion as well as baryon-decay processes involving the lightest baryon and meson octets.
$Q_a$ proves to be a strong constraint on axion interactions with strange matter.
Specifically, we obtain the first bound on the axial axion-strange-strange coupling, and the strongest existing bound on the axion-down-strange counterpart. The latter spans the range $O(10^{-1} $-$10^{-2}$) depending on thermodynamic conditions chosen to realistically represent the SN axion-emitting core. Our study uses $f_a = 10^9~\GeV$ and $m_a \ll m_{B_i, B_f, M}$, i.e. it applies to the QCD axion, which is the most difficult to constrain.
Our results suggest various additional applications, including to other compact objects and to axion-like particles with $m_a$ comparable to the external states' masses. Importantly, we would like to further investigate the dependence of our results on the SN dynamics as well as on the EoS, in particular the predicted particle abundances, along the lines discussed in the introduction. 
As a first step, one could test the effect of changing the interaction potentials that give rise to the different effective masses (and thus particle abundances), using justified ranges.
The ultimate goal is, however, a full-fledged SN simulation that includes axion losses in the dynamics, similar to neutrino losses. This simulation should also allow for the exploration of the spatial dependence (radial, and beyond) of thermodynamical quantities, of the effects of different nuclear-matter EoS and of baryon-meson couplings. For the purely nucleonic case, see e.g. Refs.~\cite{Fischer:2021jfm,Mori:2021pcv,Betranhandy:2022bvr,Mori:2023mjw}.

\noi {\bf Acknowledgments---} We warmly thank Pierluca Carenza, Alessandro Lella and Robert Ziegler for important feedback on their results. We also acknowledge useful discussions with Andrea Caputo, Alexandre Carvunis and Kiwoon Choi. Finally, we are grateful to Yann Mambrini, Pierre Salati and Guillaume Voisin for a critical reading of the manuscript.
This work has received funding from the French ANR, under contracts ANR-19-CE31-0016 (`GammaRare'), ANR-22-CE31-0001-01 (`GW-HNS') and ANR-23-CE31-0018 (`InvISYble'), that we gratefully acknowledge.
HS is supported by the Deutsche Forschungsgemeinschaft under Germany Excellence Strategy~--~EXC 2121 ``Quantum Universe''~--~390833306.

\bibliography{bibliography}

\onecolumngrid

\newpage

\begin{center}
\large \bf Supplemental Material
\end{center}

\appendix
\renewcommand\appendixname{}

\renewcommand{\thesubsection}{\thesection\arabic{subsection}}

\section{List of allowed processes} \label{app:processes}

\noi In Tables \ref{tab:Nimba}-\ref{tab:Ximba} we collect all possible processes of the kind $\Bimba$, with $B_{i,f}$ initial- or final-state octet baryons, $M$ octet mesons, and $a$ the axion. Processes with a trailing $^{\ddag; \dag; \perp}$ do not receive a contribution from, respectively the fully local, 4-leg vertex in \reffig{fig:BiM}; axion-strahlung from the meson propagator; axion-strahlung from baryon propagators (see \reffig{fig:BiM}).

The list of allowed decays of the kind $\Biba$ includes
$\Lambda \to n \, a$,~
$\Sigma^{+} \to p \, a$,~
$\Sigma^{0} \to n \, a$,~
$\Sigma^{0} \to \Lambda \, a$,~
$\Xi^{0} \to \Lambda \, a$,~
$\Xi^{0} \to \Sigma^{0} \, a$,~
$\Xi^{-} \to \Sigma^{-} \, a$.

\begin{table}[h!]
\begin{center}

\def\arraystretch{1.4}
\footnotesize
\rowcolors{1}{lightgray}{white}

\begin{tabular}{p{0.21\linewidth} | p{0.79\linewidth}}
{\bf Class} & {\bf Processes} \\
\hline

$N \, \pi \to N \, a$ & \footnotesize $n \, \pi^0 \to n \, a^{~\ddag~\dag}$, $n \, \pi^+ \to p \, a^{}$, $p \, \pi^- \to n \, a^{}$, $p \, \pi^0 \to p \, a^{~\ddag~\dag}$,  \\
$N \, \pi \to \Lambda \, a$ & \footnotesize $n \, \pi^0 \to \Lambda \, a^{}$, $p \, \pi^- \to \Lambda \, a^{}$,  \\
$N \, \pi \to \Sigma \, a$ & \footnotesize $n \, \pi^- \to \Sigma^- \, a^{}$, $n \, \pi^0 \to \Sigma^0 \, a^{}$, $n \, \pi^+ \to \Sigma^+ \, a^{~\ddag~\dag}$, $p \, \pi^- \to \Sigma^0 \, a^{}$, $p \, \pi^0 \to \Sigma^+ \, a^{}$,  \\
$N \, K \to N \, a$ & \footnotesize $n \, K^0 \to n \, a^{}$, $n \, \bar K^0 \to n \, a^{}$, $n \, K^+ \to p \, a^{}$, $p \, K^- \to n \, a^{}$, $p \, K^0 \to p \, a^{}$, $p \, \bar K^0 \to p \, a^{}$,  \\
$N \, K \to \Lambda \, a$ & \footnotesize $n \, \bar K^0 \to \Lambda \, a^{}$, $p \, K^- \to \Lambda \, a^{}$,  \\
$N \, K \to \Sigma \, a$ & \footnotesize $n \, K^- \to \Sigma^- \, a^{}$, $n \, \bar K^0 \to \Sigma^0 \, a^{}$, $p \, K^- \to \Sigma^0 \, a^{}$, $p \, \bar K^0 \to \Sigma^+ \, a^{}$,  \\
$N \, K \to \Xi \, a$ & \footnotesize $n \, K^- \to \Xi^- \, a^{~\ddag~\dag}$, $n \, \bar K^0 \to \Xi^0 \, a^{~\ddag~\dag}$, $p \, K^- \to \Xi^0 \, a^{~\ddag~\dag}$,  \\
$N \, \eta \to N \, a$ & \footnotesize $n \, \eta \to n \, a^{~\ddag~\dag}$, $p \, \eta \to p \, a^{~\ddag~\dag}$,  \\
$N \, \eta \to \Lambda \, a$ & \footnotesize $n \, \eta \to \Lambda \, a^{}$,  \\
$N \, \eta \to \Sigma \, a$ & \footnotesize $n \, \eta \to \Sigma^0 \, a^{}$, $p \, \eta \to \Sigma^+ \, a^{}$,  \\

\end{tabular}
\end{center}
\caption{$\Bimba$ processes, with $B_{i} = N$.}
\label{tab:Nimba}
\end{table}

\begin{table}[h!]
\begin{center}

\def\arraystretch{1.4}
\footnotesize
\rowcolors{1}{white}{lightgray}

\begin{tabular}{p{0.21\linewidth} | p{0.79\linewidth}}

$\Lambda \, \pi \to N \, a$ & \footnotesize $\Lambda \, \pi^0 \to n \, a^{}$, $\Lambda \, \pi^+ \to p \, a^{}$,  \\
$\Lambda \, \pi \to \Lambda \, a$ & \footnotesize $\Lambda \, \pi^0 \to \Lambda \, a^{~\ddag~\dag}$,  \\
$\Lambda \, \pi \to \Sigma \, a$ & \footnotesize $\Lambda \, \pi^- \to \Sigma^- \, a^{}$, $\Lambda \, \pi^0 \to \Sigma^0 \, a^{~\ddag~\dag}$, $\Lambda \, \pi^+ \to \Sigma^+ \, a^{}$,  \\
$\Lambda \, \pi \to \Xi \, a$ & \footnotesize $\Lambda \, \pi^- \to \Xi^- \, a^{}$, $\Lambda \, \pi^0 \to \Xi^0 \, a^{}$,  \\
$\Lambda \, K \to N \, a$ & \footnotesize $\Lambda \, K^0 \to n \, a^{}$, $\Lambda \, K^+ \to p \, a^{}$,  \\
$\Lambda \, K \to \Lambda \, a$ & \footnotesize $\Lambda \, K^0 \to \Lambda \, a^{}$, $\Lambda \, \bar K^0 \to \Lambda \, a^{}$,  \\
$\Lambda \, K \to \Sigma \, a$ & \footnotesize $\Lambda \, K^- \to \Sigma^- \, a^{}$, $\Lambda \, K^0 \to \Sigma^0 \, a^{}$, $\Lambda \, \bar K^0 \to \Sigma^0 \, a^{}$, $\Lambda \, K^+ \to \Sigma^+ \, a^{}$,  \\
$\Lambda \, K \to \Xi \, a$ & \footnotesize $\Lambda \, K^- \to \Xi^- \, a^{}$, $\Lambda \, \bar K^0 \to \Xi^0 \, a^{}$,  \\
$\Lambda \, \eta \to N \, a$ & \footnotesize $\Lambda \, \eta \to n \, a^{}$,  \\
$\Lambda \, \eta \to \Lambda \, a$ & \footnotesize $\Lambda \, \eta \to \Lambda \, a^{~\ddag~\dag}$,  \\
$\Lambda \, \eta \to \Sigma \, a$ & \footnotesize $\Lambda \, \eta \to \Sigma^0 \, a^{~\ddag~\dag}$,  \\
$\Lambda \, \eta \to \Xi \, a$ & \footnotesize $\Lambda \, \eta \to \Xi^0 \, a^{}$,  \\

\end{tabular}
\end{center}
\caption{$\Bimba$ processes, with $B_{i} = \Lambda$.}
\label{tab:Limba}
\end{table}

\begin{table}[h!]
\begin{center}

\def\arraystretch{1.4}
\footnotesize
\rowcolors{1}{lightgray}{white}

\begin{tabular}{p{0.21\linewidth} | p{0.79\linewidth}}
{\bf Class} & {\bf Processes} \\
\hline

$\Sigma \, \pi \to N \, a$ & \footnotesize $\Sigma^- \, \pi^+ \to n \, a^{}$, $\Sigma^0 \, \pi^0 \to n \, a^{}$, $\Sigma^0 \, \pi^+ \to p \, a^{}$, $\Sigma^+ \, \pi^- \to n \, a^{~\ddag~\dag}$, $\Sigma^+ \, \pi^0 \to p \, a^{}$,  \\
$\Sigma \, \pi \to \Lambda \, a$ & \footnotesize $\Sigma^- \, \pi^+ \to \Lambda \, a^{}$, $\Sigma^0 \, \pi^0 \to \Lambda \, a^{~\ddag~\dag}$, $\Sigma^+ \, \pi^- \to \Lambda \, a^{}$,  \\
$\Sigma \, \pi \to \Sigma \, a$ & \footnotesize $\Sigma^- \, \pi^0 \to \Sigma^- \, a^{~\ddag~\dag}$, $\Sigma^- \, \pi^+ \to \Sigma^0 \, a^{}$, $\Sigma^0 \, \pi^- \to \Sigma^- \, a^{}$, $\Sigma^0 \, \pi^0 \to \Sigma^0 \, a^{~\ddag~\dag}$, $\Sigma^0 \, \pi^+ \to \Sigma^+ \, a^{}$, $\Sigma^+ \, \pi^- \to \Sigma^0 \, a^{}$, $\Sigma^+ \, \pi^0 \to \Sigma^+ \, a^{~\ddag~\dag}$,  \\
$\Sigma \, \pi \to \Xi \, a$ & \footnotesize $\Sigma^- \, \pi^0 \to \Xi^- \, a^{}$, $\Sigma^- \, \pi^+ \to \Xi^0 \, a^{~\ddag~\dag}$, $\Sigma^0 \, \pi^- \to \Xi^- \, a^{}$, $\Sigma^0 \, \pi^0 \to \Xi^0 \, a^{}$, $\Sigma^+ \, \pi^- \to \Xi^0 \, a^{}$,  \\
$\Sigma \, K \to N \, a$ & \footnotesize $\Sigma^- \, K^+ \to n \, a^{}$, $\Sigma^0 \, K^0 \to n \, a^{}$, $\Sigma^0 \, K^+ \to p \, a^{}$, $\Sigma^+ \, K^0 \to p \, a^{}$,  \\
$\Sigma \, K \to \Lambda \, a$ & \footnotesize $\Sigma^- \, K^+ \to \Lambda \, a^{}$, $\Sigma^0 \, K^0 \to \Lambda \, a^{}$, $\Sigma^0 \, \bar K^0 \to \Lambda \, a^{}$, $\Sigma^+ \, K^- \to \Lambda \, a^{}$,  \\
$\Sigma \, K \to \Sigma \, a$ & \footnotesize $\Sigma^- \, K^0 \to \Sigma^- \, a^{}$, $\Sigma^- \, \bar K^0 \to \Sigma^- \, a^{}$, $\Sigma^- \, K^+ \to \Sigma^0 \, a^{}$, $\Sigma^0 \, K^- \to \Sigma^- \, a^{}$, $\Sigma^0 \, K^0 \to \Sigma^0 \, a^{}$, $\Sigma^0 \, \bar K^0 \to \Sigma^0 \, a^{}$, $\Sigma^0 \, K^+ \to \Sigma^+ \, a^{}$, $\Sigma^+ \, K^- \to \Sigma^0 \, a^{}$, $\Sigma^+ \, K^0 \to \Sigma^+ \, a^{}$, $\Sigma^+ \, \bar K^0 \to \Sigma^+ \, a^{}$,  \\
$\Sigma \, K \to \Xi \, a$ & \footnotesize $\Sigma^- \, \bar K^0 \to \Xi^- \, a^{}$, $\Sigma^0 \, K^- \to \Xi^- \, a^{}$, $\Sigma^0 \, \bar K^0 \to \Xi^0 \, a^{}$, $\Sigma^+ \, K^- \to \Xi^0 \, a^{}$,  \\
$\Sigma \, \eta \to N \, a$ & \footnotesize $\Sigma^0 \, \eta \to n \, a^{}$, $\Sigma^+ \, \eta \to p \, a^{}$,  \\
$\Sigma \, \eta \to \Lambda \, a$ & \footnotesize $\Sigma^0 \, \eta \to \Lambda \, a^{~\ddag~\dag}$,  \\
$\Sigma \, \eta \to \Sigma \, a$ & \footnotesize $\Sigma^- \, \eta \to \Sigma^- \, a^{~\ddag~\dag}$, $\Sigma^0 \, \eta \to \Sigma^0 \, a^{~\ddag~\dag}$, $\Sigma^+ \, \eta \to \Sigma^+ \, a^{~\ddag~\dag}$,  \\
$\Sigma \, \eta \to \Xi \, a$ & \footnotesize $\Sigma^- \, \eta \to \Xi^- \, a^{}$, $\Sigma^0 \, \eta \to \Xi^0 \, a^{}$,  \\

\end{tabular}
\end{center}
\caption{$\Bimba$ processes, with $B_{i} = \Sigma$.}
\label{tab:Simba}
\end{table}
\begin{table}[h!]
\begin{center}

\def\arraystretch{1.4}
\footnotesize
\rowcolors{1}{white}{lightgray}

\begin{tabular}{p{0.21\linewidth} | p{0.79\linewidth}}

$\Xi \, \pi \to \Lambda \, a$ & \footnotesize $\Xi^- \, \pi^+ \to \Lambda \, a^{}$, $\Xi^0 \, \pi^0 \to \Lambda \, a^{}$,  \\
$\Xi \, \pi \to \Sigma \, a$ & \footnotesize $\Xi^- \, \pi^0 \to \Sigma^- \, a^{}$, $\Xi^- \, \pi^+ \to \Sigma^0 \, a^{}$, $\Xi^0 \, \pi^- \to \Sigma^- \, a^{~\ddag~\dag}$, $\Xi^0 \, \pi^0 \to \Sigma^0 \, a^{}$, $\Xi^0 \, \pi^+ \to \Sigma^+ \, a^{}$,  \\
$\Xi \, \pi \to \Xi \, a$ & \footnotesize $\Xi^- \, \pi^0 \to \Xi^- \, a^{~\ddag~\dag}$, $\Xi^- \, \pi^+ \to \Xi^0 \, a^{}$, $\Xi^0 \, \pi^- \to \Xi^- \, a^{}$, $\Xi^0 \, \pi^0 \to \Xi^0 \, a^{~\ddag~\dag}$,  \\
$\Xi \, K \to N \, a$ & \footnotesize $\Xi^- \, K^+ \to n \, a^{~\ddag~\dag}$, $\Xi^0 \, K^0 \to n \, a^{~\ddag~\dag}$, $\Xi^0 \, K^+ \to p \, a^{~\ddag~\dag}$,  \\
$\Xi \, K \to \Lambda \, a$ & \footnotesize $\Xi^- \, K^+ \to \Lambda \, a^{}$, $\Xi^0 \, K^0 \to \Lambda \, a^{}$,  \\
$\Xi \, K \to \Sigma \, a$ & \footnotesize $\Xi^- \, K^0 \to \Sigma^- \, a^{}$, $\Xi^- \, K^+ \to \Sigma^0 \, a^{}$, $\Xi^0 \, K^0 \to \Sigma^0 \, a^{}$, $\Xi^0 \, K^+ \to \Sigma^+ \, a^{}$,  \\
$\Xi \, K \to \Xi \, a$ & \footnotesize $\Xi^- \, K^0 \to \Xi^- \, a^{}$, $\Xi^- \, \bar K^0 \to \Xi^- \, a^{~\perp}$, $\Xi^- \, K^+ \to \Xi^0 \, a^{}$, $\Xi^0 \, K^- \to \Xi^- \, a^{}$, $\Xi^0 \, K^0 \to \Xi^0 \, a^{}$, $\Xi^0 \, \bar K^0 \to \Xi^0 \, a^{}$,  \\
$\Xi \, \eta \to \Lambda \, a$ & \footnotesize $\Xi^0 \, \eta \to \Lambda \, a^{}$,  \\
$\Xi \, \eta \to \Sigma \, a$ & \footnotesize $\Xi^- \, \eta \to \Sigma^- \, a^{}$, $\Xi^0 \, \eta \to \Sigma^0 \, a^{}$,  \\
$\Xi \, \eta \to \Xi \, a$ & \footnotesize $\Xi^- \, \eta \to \Xi^- \, a^{~\ddag~\dag}$, $\Xi^0 \, \eta \to \Xi^0 \, a^{~\ddag~\dag}$,  \\

\end{tabular}
\end{center}
\caption{$\Bimba$ processes, with $B_{i} = \Xi$.}
\label{tab:Ximba}
\end{table}

\section{Axion-hadron interactions: explicit formul\ae} \label{app:currents}

\noi The $\mc J^{b \mu}_{L,R}$ currents in \refeq{eq:LaUB} may be decomposed into purely mesonic and baryonic-plus-$n$-mesons pieces as follows
\be
\mcJ_{L,R}^{b \mu} ~=~ (\mcJ_{L,R}^{(U)})^{b \mu} + (\mcJ_{L,R}^{(B)})^{b \mu}~,
\ee
where the purely mesonic parts are given explicitly as
\be
\beal
\label{eq:JU_LR}
(\mc J^{(U)}_L)^{\mu {b}} = + i \frac{F_0^2}{2} \Tr \left[ \frac{\la^b}{2} (D^\mu U)^\dagger U \right]~,~~~~
(\mc J^{(U)}_R)^{\mu {b}} = - i \frac{F_0^2}{2} \Tr \left[ \frac{\la^b}{2} U (D^\mu U)^\dagger \right]~.
\eeal
\ee
and the baryonic-plus-$n$-mesons parts as
\be
\beal
\label{eq:JB}
(\mcJ^{(B)}_{L,R})^{b\mu} =
(\mc J^{(B, K_1)}_{L,R})^{\mu {b}} + (\mc J^{(B, K_2)}_{L,R})^{\mu {b}} + (\mc J^{(B, D)}_{L,R})^{\mu {b}} + (\mc J^{(B, F)}_{L,R})^{\mu {b}}~.
\eeal
\ee
Adhering to the notation in Ref.~\cite{\scherer}, the first two terms on the r.h.s. come from the kinetic terms $i \Tr \left( \bar B  \slashed \partial B \right)$ and $i \Tr \left( \bar B [ \slashed \Gamma, B ] \right)$, respectively, and the last two terms from the terms proportional to the $D$ and $F$ couplings, respectively. The terms on the r.h.s. of \refeq{eq:JB} are given explicitly as
\bea
&&(\mc J^{(B, K_1)}_{L,R})^{\mu {b}} =
\frac{1}{4} \Tr \Bigl( \bar B \gamma^\mu \left[\lambda^b \mp \frac{f^{\lambda \phi b}}{2 F_0}, B \right] \Bigl)~,\\[0.3cm]
&&(\mc J^{(B, K_2)}_{L})^{\mu {b}} = 
-\frac{1}{8} \Tr \left( \bar B \gamma^\mu \left[\lambda^b - \frac{f^{\lambda \phi b}}{2 F_0} - u \left( \lambda^b - \frac{f^{\lambda \phi b}}{2 F_0} \right) u^\dagger, B \right] \right)~,\\
&&(\mc J^{(B, K_2)}_{R})^{\mu {b}} =
-\frac{1}{8} \Tr \left( \bar B \gamma^\mu \left[\lambda^b + \frac{f^{\lambda \phi b}}{2 F_0} - u^\dagger \left( \lambda^b + \frac{f^{\lambda \phi b}}{2 F_0} \right) u, B \right] \right)~,\\[0.3cm]
&&(\mc J^{(B, D)}_{L})^{\mu {b}} =
+\frac{D}{8} \Tr \left( \bar B \gamma^\mu \gamma_5\left\{\lambda^b - \frac{f^{\lambda \phi b}}{2 F_0} + u \left( \lambda^b - \frac{f^{\lambda \phi b}}{2 F_0} \right) u^\dagger, B \right\} \right)~,\\
&&(\mc J^{(B, D)}_{R})^{\mu {b}} =
-\frac{D}{8} \Tr \left( \bar B \gamma^\mu \gamma_5\left\{\lambda^b + \frac{f^{\lambda \phi b}}{2 F_0} + u^\dagger \left( \lambda^b + \frac{f^{\lambda \phi b}}{2 F_0} \right) u, B \right\} \right)~,
\eea
and
$
(\mc J^{(B, F)}_{L,R})^{\mu {b}} = (\mc J^{(B, D)}_{L,R})^{\mu {b}}|_{ \{,\} \rightarrow [,]~;~D \rightarrow F}
$.

We adhere to the notation in Ref.~\cite{\scherer}, in particular we define the mesonic field $U = \exp(i \phi / F_0)$, with covariant derivative $D_\mu U ~=~ \de_\mu U - i \frac{\de_\mu a}{f_a} \left( \hkR{} U - U \hkL{} \right) - i e A_\mu \left[ Q, U \right]$ and $Q = \diag(Q_u, Q_d, Q_s)$. We normalize the 3-flavour pion-field matrix $\phi = \phi^a \lambda^a$ (where $\lambda^a$ are the Gell-Mann matrices), so that e.g. the (1,1) entry is $\pi_0 + \eta_0/\sqrt3$, and $F_0 \approx 93$~MeV. We assume the chiral-symmetry transformation to be $U \to R U L^\dagger$. Also, we use $u = \sqrt U$, and the shortcut notation $f^{X Y c} = f^{a b c} X^a Y^b$, with $f$ the $SU(3)$ structure constants, and $a,b,c$ SU(3) indices in the adjoint representation.
Again, we normalize the hyperon field $B = B^a \lambda^a$ as in~\cite{\scherer}, such that e.g. the (1,1) entry is $\Sigma^0 / \sqrt2 + \Lambda / \sqrt6$ and we take $D = 0.81$ and $F = 0.44$ \cite{Vonk:2021sit}.

In the construction of the above currents, we neglect the contribution from the SM chiral Lagrangian mediating weak processes~\cite{Cronin:1967jq,Kambor:1989tz}, augmented with an axion.
These contributions are---for generic $\hkL{}$ and $\hkR{}$ complying with existing constraints---suppressed by a relative factor of $G_F F_0^2 \sim 10^{-7}$ with respect to $\DS1$ arising from \refeq{eq:LaUB}.

Lastly, we are not concerned about the possibly poor convergence of the baryonic chiral Lagrangian \cite{GrillidiCortona:2015jxo,Notari:2022ffe}. We do not expect such issue to impact our conclusions, e.g. to grossly affect the $Q_a$ orders of magnitude found.

\section{Technical details on Figure~\ref{fig:matrix_plots_DD2Y_Sag} and Comparison plots} \label{app:Sfl_and_SFmod_plots}

\noi As explained in the main text, Fig.~\ref{fig:matrix_plots_DD2Y_Sag} shows distributions for the $\bk$ couplings $(\vb*{k}_{A})_{11}$, $(\vb*{k}_{A})_{22}$, $(\vb*{k}_{A})_{33}$, $|(\vb*{k}_{A})_{23}|$ projected onto $1$-parameter spaces (histograms on the diagonal panels) or $2$-parameter spaces (panels off-the-diagonal). For each panel, the parameters not shown can take on any value, i.e. they are ``marginalized over''. The $\bk$ couplings are free numbers to start with. The initial distribution from each $\bk$ coupling is built by floating uniform random values in the range $[0, 10]$. We sampled values till collecting $N = 10^6$ instances that fulfil two sets of constraints:
\begin{itemize}
\item the constraint on $(\vb*{k}_{A})_{11}$ and $(\vb*{k}_{A})_{22}$ (two reals) from isolated-Neutron-Star cooling \cite{Buschmann:2021juv};
\item the constraints on $(\vb*{k}_{V,A})_{23}$ (two complex numbers) from $K$-physics observables.
\end{itemize} The $\bk$ couplings distributions are required to fulfil the ensemble of these constraints (described in detail in the top paragraph on page 3, left column), before imposing the Supernova constraint $Q_a < Q_\nu$~\cite{Raffelt:1990yz,Burrows:1988ah} that forms the object of our study. The ensemble of $\bk$ couplings that pass this last constraint form the output distributions shown in \reffig{fig:matrix_plots_DD2Y_Sag}.

\begin{figure*}[th!]
  \centering
  \includegraphics[width=\textwidth]{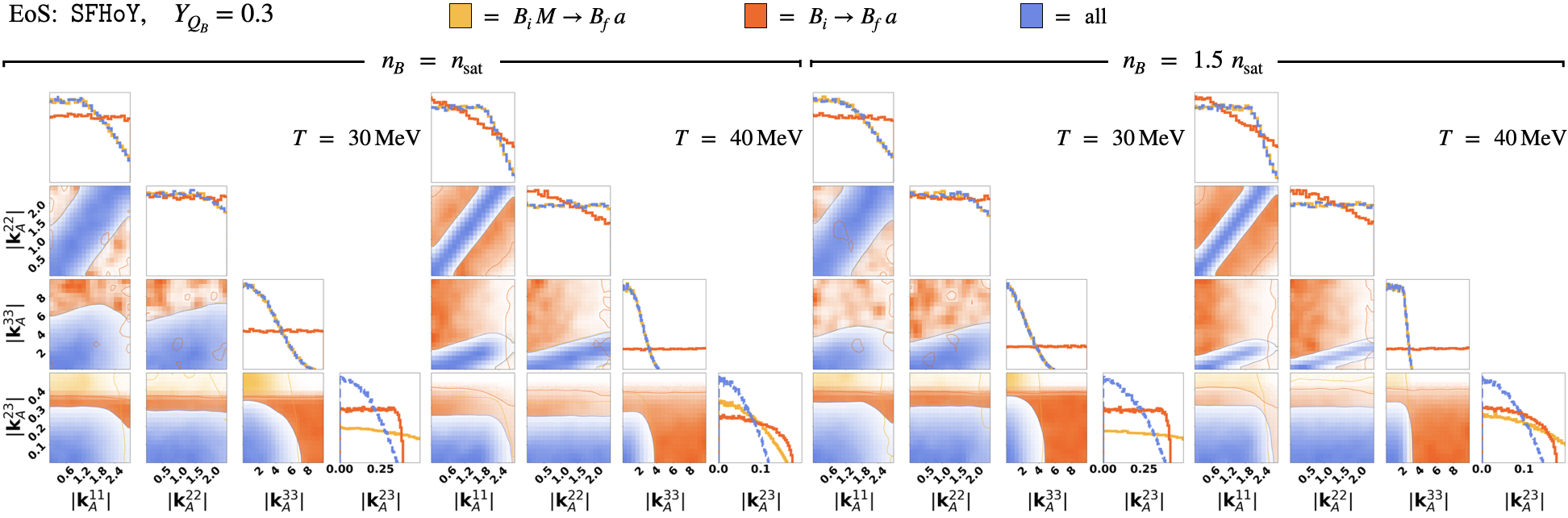}
  \caption{Same as Fig.~\ref{fig:matrix_plots_DD2Y_Sag}, but for using the \SFmod EoS.}
  \label{fig:matrix_plots_SFHoY}
\end{figure*}

We verified the stability of these distributions with respect to varying $N$ (see definition above). Specifically, we increased $N$ until the distributions for each parameter stabilized in shape. Once we observed no further variation, we fixed the number of points accordingly. The chosen $N = 10^6$ is thus the largest value we tested.
For completeness, we also show in \reffig{fig:matrix_plots_SFHoY} the \SFmod counterparts of Fig.~\ref{fig:matrix_plots_DD2Y_Sag}.

\section{Details on SN particle fractions} \label{app:snmodel}
\begin{table*}[ht]
  \caption{Particle fractions in SN matter for the different thermodynamic conditions considered, as specified in the first two columns, and $Y_{Q_B} = 0.3$. Only particle fractions $> 10^{-5}$ are listed.}
  \label{tab:snmodel_fractions}
  \centering
  \begin{tabularx}{1.0\columnwidth}{
>{\centering\arraybackslash}p{0.037\columnwidth}
>{\centering\arraybackslash}p{0.049\columnwidth}
>{\centering\arraybackslash}p{0.059\columnwidth}
>{\centering\arraybackslash}p{0.059\columnwidth}
>{\centering\arraybackslash}p{0.059\columnwidth}
>{\centering\arraybackslash}p{0.059\columnwidth}
>{\centering\arraybackslash}p{0.074\columnwidth}
>{\centering\arraybackslash}p{0.074\columnwidth}
>{\centering\arraybackslash}p{0.074\columnwidth}
>{\centering\arraybackslash}p{0.074\columnwidth}
>{\centering\arraybackslash}p{0.074\columnwidth}
>{\centering\arraybackslash}p{0.059\columnwidth}
>{\centering\arraybackslash}p{0.074\columnwidth}
>{\centering\arraybackslash}p{0.074\columnwidth}
    }
  \toprule
  & $T$ & $Y_e$& $Y_n$ &$Y_p$& $Y_{\Lambda}$& $Y_{\Sigma^-}$& $Y_{\Sigma^0}$& $Y_{\Sigma^+}$& $Y_{\Xi^-}$& $Y_{\Xi^0}$& $Y_{\pi^-}$& $Y_{\pi^0}$& $Y_{\pi^+}$ \\
  & (MeV) & & & \\
  \midrule
  \addlinespace
 \multirow{2}{*}{\rotatebox{90}{\begin{tabular}{c}{%
 \DDmod}\\{\small $n_B = n_{\sat}$}\end{tabular}}} %
 & 30 & 0.298 & 0.696 & 0.300 & 0.004 & $2\times 10^{-4}$ & $5\times 10^{-5}$ & $1\times 10^{-5}$ & $3\times 10^{-5}$ & $1\times 10^{-5}$ & 0.002 & $2\times 10^{-4}$ & $2\times 10^{-5}$ \\
  \addlinespace
  \addlinespace
  \\
 & 40 & 0.295 & 0.680 & 0.302 & 0.014 & 0.002 & 6$\times 10^{-4}$ & $2\times 10^{-4}$ & 5$\times 10^{-4}$ & 2$\times 10^{-4}$ & 0.006 & 0.001 & 2$\times 10^{-4}$ \\
  \addlinespace
  \midrule
  \addlinespace
 \multirow{2}{*}{\rotatebox{90}{\begin{tabular}{c}{%
 \SFmod}\\{\small $n_B = n_{\sat}$}\end{tabular}}} %
 & 30 & 0.298 & 0.697 & 0.300 & 0.003 & $1\times 10^{-4}$ & $4\times 10^{-5}$ & $1\times 10^{-5}$ & $2\times 10^{-5}$ &  & 0.002 & $2\times 10^{-4}$ & $2\times 10^{-5}$ \\
  \addlinespace
  \addlinespace
  \\
 & 40 & 0.294 & 0.686 & 0.301 & 0.011 & 0.001 & 5$\times 10^{-4}$ & 2$\times 10^{-4}$ & 3$\times 10^{-4}$ & 1$\times 10^{-4}$ & 0.006 & 0.001 & 2$\times 10^{-4}$ \\
  \addlinespace
  \midrule
  \addlinespace
  \addlinespace
 \multirow{2}{*}{\rotatebox{90}{\begin{tabular}{c}{%
 \DDmod}\\{\small $n_B = 1.5\, n_{\sat}$}\end{tabular}\hspace{-0.3cm}}} %
 & 30 & 0.299 & 0.690 & 0.301 & 0.009 & 5$\times 10^{-4}$ & 8$\times 10^{-5}$ & 1$\times 10^{-5}$ & 1$\times 10^{-4}$ & 3$\times 10^{-5}$ & 0.001 & 1$\times 10^{-4}$ & 1$\times 10^{-5}$ \\
  \addlinespace
  \addlinespace
  \\
 & 40 & 0.296 & 0.667 &	0.304 & 0.023 & 0.003 & 8$\times 10^{-4}$ & 2$\times 10^{-4}$ & 0.001 & 3$\times 10^{-4}$ & 0.004 & 8$\times 10^{-4}$ & 1$\times 10^{-4}$ \\
  \addlinespace
  \addlinespace
  \midrule
  \addlinespace
  \addlinespace
 \multirow{2}{*}{\rotatebox{90}{\begin{tabular}{c}{%
 \SFmod}\\{\small $n_B = 1.5\, n_{\sat}$}\end{tabular}\hspace{-0.3cm}}} %
 & 30 & 0.299 & 0.697 & 0.300 & 0.003 & 8$\times 10^{-5}$ & 2$\times 10^{-5}$ & & 2$\times 10^{-5}$ & & 0.001 & 1$\times 10^{-4}$ & 1$\times 10^{-5}$ \\
  \addlinespace
  \addlinespace
  \\
 & 40 & 0.295 & 0.686 & 0.301 & 0.011 & 9$\times 10^{-4}$ & 3$\times 10^{-4}$ & 1 $\times 10^{-4}$ & 4$\times 10^{-4}$ & 1$\times 10^{-4}$ & 0.005 & 8$\times 10^{-4}$ & 1$\times 10^{-4}$ \\
  \addlinespace
  \addlinespace
  \bottomrule
  \end{tabularx}
\end{table*}

\noi Thermodynamic quantities and effective masses and chemical potentials for our two considered EoS models, \DDmod and \SFmod, are available for large ranges of $T, n_B$ and $Y_{Q_B}$ in tabulated form from the \textsc{CompOSE} database~\cite{Typel:2013rza,CompOSECoreTeam:2022ddl}.
In Table~\ref{tab:snmodel_fractions} we show the full list of particle fractions for the different thermodynamic conditions considered. As the table shows, the choice $Y_{Q_B} = 0.3$ yields $Y_e \approx 0.3$, consistently with the value used in Ref.~\cite{\carenza}.

\end{document}